\documentclass[12pt]{article}
\usepackage{a4}\usepackage{epsf}

\newcommand{\be}{\begin{equation}}
\newcommand{\ee}{\end{equation}}
\newcommand{\bea}{\begin{eqnarray}}
\newcommand{\eea}{\end{eqnarray}}
\newcommand{\beq}{\begin{eqnarray}}
\newcommand{\eeq}{\end{eqnarray}}
\newcommand{\beao}{\begin{eqnarray*}}
\newcommand{\eeao}{\end{eqnarray*}}
\newcommand{\nn}{\nonumber}
\newcommand{\pa}{\partial}
\newcommand{\la}{\lambda}

\newcommand{\rinf}{{\raisebox{-7pt}{$\sim$}\atop  {\mbox{$\scriptstyle   r\to\infty$}}} \ }
\newcommand{\Ref}[1]{(\ref{#1})}
\newcommand{\al}{{\alpha}}

\newcommand{\E}{{\cal E}}
\newcommand{\Eq}{{\cal E_{\rm quant}}}
\newcommand{\Ef}{{\cal E^{\rm f}}}
\newcommand{\Eas}{{\cal E^{\rm as}}}
\newcommand{\Efn}{{\cal E}^{\rm f}_0}
\newcommand{\Easn}{{\cal E}^{\rm as}_0}
\newcommand{\uv}{ultraviolet }
\begin{document}
\title{On the Vacuum energy of a Color Magnetic Vortex} 
\author{
{\sc M. Bordag}\thanks{e-mail: Michael.Bordag@itp.uni-leipzig.de} \\
\small  University of Leipzig, Institute for Theoretical Physics\\
\small  Augustusplatz 10/11, 04109 Leipzig, Germany}
\maketitle
\begin{abstract}
 We calculate the one loop gluon vacuum energy in the background of a
 color magnetic vortex for SU(2) and SU(3). We use zeta
 functional regularization to obtain analytic expressions suitable for
 numerical treatment. The momentum integration is turned to the
 imaginary axis and fast converging sums/integrals are obtained. We
 investigate numerically a number of profiles of the background. In
 each case the vacuum energy turns out to be positive increasing in this
 way the complete energy and making the vortex configuration less
 stable. In this problem bound states (tachyonic modes) are present
 for all investigated profiles making them intrinsically unstable.
\end{abstract}
\section{Introduction}\label{Sec1}
Color magnetic vortices or center-of-group vortices play at present an
important role in the discussion of possible mechanisms for
confinement in QCD
\cite{'tHooft:1978hy,DelDebbio:1997mh,Engelhardt:1999wr}.  There are
recent investigations of vortex like configurations in QED
\cite{Langfeld:2002vy} and in QCD \cite{Diakonov:2002bx}. Given the
actual interest in this topic we consider in the present paper the
case of Abelian color magnetic vortices with arbitrary profile for
SU(2) and SU(3) gluodynamics. Following the technique developed in
\cite{Bordag:1998tg} for a magnetic flux tube and earlier in
\cite{Bordag:1996fv} for a spherically symmetric background field we
write down analytic expressions for the vacuum energy well suited for
numerical investigation.  Using zeta functional regularization we
express the vacuum energy in terms of the Jost function of the related
scattering problem, and by separating a piece of the asymptotic
expansion of the Jost function we obtain fast converging sums and
integrals.

At present, several calculation schemes are known. The oldest and
perhaps most intuitive one is to sum over phase shifts or, what is
equivalent, over mode densities. For recent applications see
\cite{Diakonov:2002bx} and \cite{Graham:2002fi}. This method is
numerically demanding because of the oscillatory behavior of the phase
shifts as function of the radial momentum. Another method that had
been proposed in \cite{Bordag:1996fv,Bordag:1998tg} uses the Jost
function and a momentum integration turned into the imaginary axis in
this way avoiding oscillatory integrands (for a detailed description
of these methods see also \cite{KirstenBook}). A completely different
approach is that of \cite{Gies:2001tj,Gies:2001zp} using world line
methods. Its main advantage is that it does not require separation of
variables. Recently, this method was applied to a magnetic string and
numerical results similar to those in \cite{Bordag:1998tg} have been
obtained (for a comparizon see \cite{Pasipoularides:2000gg}.

The issue of renormalization of vacuum energies has been much
discussed in the past and different schemes are in use. Finally, all
differences are settled by an appropriate normalization condition and
the many schemes are different ways to the same destination. However,
this is not always easy to see. Let us first consider theories with a
massive quantum field, a spinor for instance.  Here the mass of the
field provides the opportunity to formulate the 'large mass'
normalization condition stating the vacuum energy must vanish if this
mass becomes large \cite{Bordag:2000f,Bordag:1999vs}. This condition
was initially formulated in connection with the Casimir effect for
massive fields \cite{Bordag:1997ma}. In \cite{BGNV} it was shown to be
equivalent to the no tadpole normalization condition known in quantum
field theory. This condition is directly related to the heat kernel
expansion which is equivalent to the expansion of Greens functions in
inverse powers of the mass \cite{dewi65b}. It is also evident that
this condition is in line with the charge renormalization appearing
from vacuum energy calculations in QED. Another picture we observe
with massless quantum fields. There is no such normalization condition
causing, for instance, the known problems with the vacuum energy of a
dielectric sphere \cite{Bordag:1999vs}. However, in the present paper
we are concerned with QCD and the vacuum fluctuation of the gluons.
Although they are massless, in QCD there is the commonly accepted
coupling constant renormalization involving the phenomenological
quantity $\Lambda_{QCD}$. We will use this scheme. The only technical
problem is that this scheme is frequently used together with the
Pauli-Villars regularization while we use the zeta functional one. But
the relation between both schemes can easily be written down, see
Sect. 2.

The paper is organized as follows. In the next section we collect the
general formulas for the color vortex and its vacuum energy. Following
\cite{Diakonov:2002bx} we write down the condition for the minimum of
the complete energy. In the two subsequent sections we consider in
detail the finite and the asymptotic parts of the vacuum energy for
the given configurations. In Sec. 5 the numerical results are
represented. Sec. 6 contains a discussion and some technical details
are given in the Appendix.

We hope to have represented all formulas in such detail, or given the
corresponding references, that the reader is able to repeat the
calculations.


\section{The color magnetic vortex and its vacuum energy}\label{Sec2}
We consider a non Abelian gauge field with standard action
\be\label{action}S=\frac{1}{4g^2}\int d^4x \ \left(F_{\mu\nu}^a(x)\right)^2.
\ee
The corresponding potential, $A_{\mu}^a(x)$, is expanded around a
background, $A_{\mu}^{a({\rm bg})}(x)$,
\be\label{field1}A_{\mu}^a(x)=
A_{\mu}^{a({\rm bg})}(x)+A_{\mu}^{a({\rm qu})}(x),
\ee
where $A_{\mu}^{a({\rm qu})}(x)$ are the quantum fluctuations. The
background is a center-of-group vortex, i.e., it is an Abelian static
and cylindrically symmetric configuration. The corresponding color
magnetic field is parallel to the third axis in both, color and space,
\be\label{bgr}\vec{B}^a(x)=\delta^{a,3} \ \vec{e}_{z} \ \frac{\mu'(r)}{r},
\ee
where $\mu(r)$ is a profile function, $\vec{e}_{z}$ is the unit vector
parallel to the z-axis and $\delta^{a,3}$ is the Kronecker symbol. The
classical energy of this field meant as energy density per unit length
with respect to the z-axis is
\be\label{Eclass}\E_{\rm
  class}=\frac{\pi}{g^2}\int_0^\infty\frac{dr}{r} \ \mu'(r)^2.
\ee
The profile function will be chosen vanish at $r=0$.  Its first
derivative must vanish too other\-wise the classical energy is not
finite. For $r\to\infty$ it must tend to a constant which determines
the flux. We consider examples with $\mu(\infty)=1$.

The fluctuation fields $A_{\mu}^{a({\rm qu})}(x)$ are quantized in a
standard way. We use the background gauge condition as in
\cite{Diakonov:2002bx} and have gluon and ghost contributions. The
vacuum energy due to these fields is
\be\label{eqallg}\Eq=\E_{\rm quant}^{\rm gluon}+\E_{\rm quant}^{\rm ghost}.
\ee
In general, a vacuum energy is given by
\be\label{eqs}\Eq=\pm\frac12\sum_{(n)}e_{(n)}^{1-2s},
\ee
where $e_{(n)}$ are the corresponding one particle energies and the
sum over $(n)$ is over all quantum numbers present for the given
field. The parameter $s$ is the regularization parameter of the
intermediate zeta functional regularization which we use.  We have to
take $s>\frac32$ in the beginning and to put $s=0$ in the end. We
ignore the arbitrary parameter $\mu$ which is usually introduced in
this scheme.  The sign is +1 for gluons and -1 for ghosts.  For both
types of particles there is a sum over color and polarization. For the
ghosts there is an additional factor of 2 because there are two
fields, $c$ and $\bar{c}$. In the considered background, variables
separate in the equations of motion for the fluctuating fields. One
can separate the dependence on the z-coordinate and the angular
dependence in the plane perpendicular to the z-axis introducing $p_z$
and the angular momentum $\nu$ ($\nu=0,\pm1,\pm2,\dots$) accordingly.
After that the one particle energies read
\be\label{en}e_{(n)}=\sqrt{p_z^2+{\lambda_{\nu,n}^{(c,\la)}}^2},
\ee
where $\lambda_{\nu,n}^{(c,\la)}$ are the eigenvalues of the residual
radial equation,
\be\label{radeq}\left(-\frac{1}{r}\frac{\pa}{\pa r}r\frac{\pa}{\pa   r}
+\frac{(\nu-\al_c\mu(r))^2}{r^2}
-\frac{2\al_c\beta_\la\mu'(r)}{r}    \right)  \phi^{(c,\la)}_{\nu,n}(r)
={\lambda_{\nu,n}^{(c,\la)}}^2         \          \phi^{(c,\la)}_{\nu,n}(r).
\ee
Here we assume the radial quantum number $n$ to be discrete, see below.
This equation emerges after diagonalization in color space. The
corresponding color eigenvalues are
\be\label{alc}\al_c=\left\{\begin{array}{ll} 
\{0,-1,1\}&\mbox{for SU(2)}\\
\{0,0,-\frac12,-\frac12,\frac12,\frac12,-1,1\}&\mbox{for SU(3)}\ .
\end{array}  \right. \  
\ee
For the ghost field we have to put $\beta_\la=0$ in
Eq.\Ref{radeq}. For the gluon the polarizations $\beta_\la$ are
\be\label{betla}\beta_\la=\{0,0,-1,1\}.
\ee
In the vacuum energy, the two gluon polarizations with $\beta_\la=0$
cancel the ghost contribution. So we are left with the two gluon
polarizations $\beta_\la=\pm1$ and no ghosts.

In the next step we integrate over $p_z$ by means of 
\[\label{pz}\int_{-\infty}^\infty \frac{dp_z}{2\pi} \
\left(p_z^2+\la^2\right)^{\frac12-s} 
=
\la^{2(1-s)}\frac{\Gamma(s-1)}{2\sqrt{\pi}\Gamma(s-\frac12)}
=
\la^{2(1-s)}\frac{C_s}{4\pi s} 
\]
with $C_s=1+s(2\ln 2-1)+\dots$ and arrive at
\be\label{eq2}\Eq=\frac{C_s}{2}\sum_{c,\la}\sum_{\nu=-\infty}^{\infty} 
\sum_{n}{\lambda_{\nu,n}^{(c,\la)}}^{2(1-s)}.
\ee
With this representation and equation \Ref{radeq} the whole problem is
reduced to a number of scalar ones. Starting from here, and for pure
technical convenience we introduce an auxiliary mass $m$ by means of
\be\label{auxmass}
{\lambda_{\nu,n}^{(c,\la)}} \to 
\left({\lambda_{\nu,n}^{(c,\la)}}^2+m^2\right)^{1/2}
\ee
with $m\to0$ before the end of calculations.

It remains to transform the discrete sum over the radial quantum
number $n$ into an integral. As the initial problem is defined on the
unbounded plane perpendicular to the z-axis this quantum number is
continuous. We make it discrete by first imposing a large 'box', i.e.,
a cylinder at some large radius $R$ with some boundary conditions. We
follow \cite{Bordag:1998tg} where this procedure is described in
detail, but we discuss all subtle specifics of the given case.
So, omitting all indices, let $\phi(r)$ be the regular solution of the
radial equation on the whole axis. It is defined as that solution
which at $r\to0$ turns into the solution for $\mu(r)=0$ which is
simply the Bessel function $J_\nu(kr)$.  For large $r$, $\phi(r)$ has
the asymptotic expansion
\be\label{defJ}\phi(r)     \rinf 
\frac12\left(  
f_\nu(k)\  H^{(2)}_\nu(kr)+ \overline{f}_\nu(k) \ H^{(1)}_\nu(kr) 
       \right).
\ee
The coefficients $f_\nu(k)$ and its complex conjugate are the Jost
functions. The Hankel functions are also solutions of the radial
equation \Ref{radeq} for $\mu(r)=0$. However, we have to take into
account $\mu(r)\rinf 1$ and should use the Hankel functions
$H^{(1,2)}_{\nu-\al_c}(kr)$ instead.  For the definition of
the Jost function, this doesn't matter because the leading term in the
asymptotic expansion of the Hankel functions for large argument
doesn't depend on the index. 

Following \cite{Bordag:1998tg}, for large $R$, the boundary condition
$\phi(R)=0$ defines the discrete eigenvalues $k\to {\lambda_n}$. This
allows to write the sum over $n$ as a contour integral encircling the
spectrum with the logarithmic derivative of $\phi(R)$ generating the
necessary poles. After that we tend $r$ to infinity and drop a
contribution which does not depend on the background (the so called
Minkowski space contribution). In the heat kernel terminology this is
to drop the coefficient $a_0$'s contribution. Finally turning the
integration contour toward the imaginary axis we arrive at the
representation
\be\label{1}\sum_n \left(\la_n^2+m^2\right)^{1-s}=
\frac{\sin \pi  s}{-\pi}    \int_m^\infty dk \ 
\left(k^2-m^2\right)^{1-s}\frac{\pa}{\pa k} \ln f_\nu(ik).
\ee
Putting these things together, we get for the vacuum energy \Ref{eq2},
\be\label{eqr}\Eq=
\frac{C_s}{-8\pi }\sum_{c,\la}\sum_{\nu=-\infty}^{\infty}  
\int_m^\infty dk \ 
\left(k^2-m^2\right)^{1-s}\frac{\pa}{\pa k} \ln f_\nu^{(c,\la)}(ik),
\ee
where now $f_\nu^{(c,\la)}(ik)$ is the Jost function corresponding to
Eq.\Ref{radeq} and we kept only first order corrections for $s\to0$.
This is a very convenient representation for the vacuum energy. The
main advantage is that the integration goes over the imaginary axis
avoiding oscillating functions. In addition we remark that bound
states, if present in the spectrum, are included in this formula
automatically\footnote{This is discussed in detail in
\cite{Bordag:1995jz} and used in numerical examples in
\cite{Bordag:1999sf}}. Bound states manifest themselves as zeros of
the Jost function on the positive imaginary axis. If the binding
energy $\kappa_i$ is smaller that the mass $m$ of the fluctuating
field the corresponding poles of the logarithmic derivative are
outside the integration region in Eq.\Ref{eqr} and the vacuum energy
is real.  In opposite case, for $m<\kappa_i$, the integration runs
over the pole (it must be bypassed on the right) and an imaginary part
appears. Using $\int \frac{dk}{k-i0}=i\pi$ we obtain
\be\label{eqim}\Im \Eq=\frac{-1}{8}\sum_{c,\la}\sum_\nu\sum_i(\kappa_i^2-m^2),
\ee
where the sums go over those values of the quantum numbers for which a
bound state is present. For instance, $i$ is limited by
$m<\kappa_i$. In our case we have $m=0$ and any bound state
contributes to the imaginary part of the vacuum energy. We note that
the imaginary part does not contain \uv divergences and so that we
could put $s=0$ here.  The real part of the vacuum energy is then
given by Eq.\Ref{eqr} taking the integration over $k$ as principal
value.

The vacuum energy given by Eq.\Ref{eqr} still contains known \uv
divergencies. Their structure can be best discussed in terms of the
heat kernel coefficients. The contribution of the coefficient $a_0$
we already dropped. Due to the gauge invariance the coefficent $a_1$
is zero. The next coefficient is $a_2$ which is here given by
\be\label{a2}a_2=\frac{-22N}{3}\pi\int \frac{dr}{r} \mu'(r)^2.
\ee
At once it is the last
coefficient which contributes to the \uv divergencies. By writing its
contribution to the vacuum energy separately (for a derivation of the
corresponding formula see \cite{Bordag:1996fv}) we define by means of
\be\label{eqd}\Eq=
\frac{-a_2}{32\pi^2}\left(\frac1s-2-2\ln\frac{m}{2}\right)+
\E_{\rm quant}^{\rm ren}
\ee
the renormalized vacuum energy $\E_{\rm quant}^{\rm ren}$. In general,
after separating the \uv divergence, the remaining finite (in the
limit of removing the regularization) part is not uniquely defined. In
the form given here it contains contributions from the heat kernel
coefficients $a_n$ with numbers larger than 2 only. Because the heat
kernel expansion provides an expansion in inverse powers, $m^{4-2n}$,
of the mass of the fluctuating field, $\E_{\rm quant}^{\rm ren}$ goes
to zero for $m\to\infty$. In problems with a massive fluctuating field
we used this property as normalization condition defining in this way
a unique renormalized vacuum energy \cite{Bordag:1997ma,Bordag:2000f},
see also \cite{Elizalde:1997hx}. For massless fields this does not
work. In the case considered in this paper the underlying theory is
QCD. Here the renormalization is done in the Pauli-Villars scheme by a
redefinition of the bare coupling constant in Eq.\Ref{action}
according to
\be\label{g}\frac{1}{g^2}=\frac{11N}{24\pi^2}\ln\frac{M}{\Lambda_{\rm QCD}},
\ee
where $M$ is the Pauli-Villars mass and $N$ is the order of the gauge
group, SU(N). We denote for a moment the regularized vacuum energy by
$\Eq(s,m)$, indicating the dependence on the regularization parameter
$s$ and on the mass $m$ explicitly.  In the Pauli-Villars scheme the
vacuum energy is given by
\be\label{eqPV}\E_{\rm quant}^{\rm PV}=\Eq(s,m)-\Eq(s,M)
\ee
for $m\to0$ and $M\to\infty$. Using the representation of the vacuum
energy given by Eq.\Ref{eqd} as a result of the heat kernel
decomposition we obtain
\bea\label{eq1PVp}\E_{\rm quant}^{\rm PV}&=&
\frac{-a_2}{32\pi^2}\left(\frac1s-2-2\ln\frac{m}{2}\right)+
\E_{\rm quant}^{\rm ren}(s,m)  \nn \\  &&
-\frac{-a_2}{32\pi^2}\left(\frac1s-2-2\ln\frac{M}{2}\right)
-\E_{\rm quant}^{\rm ren}(s,M).
\eea
Here a number of contributions proportional to $a_2$ cancel, the pole
term for instance\footnote{Starting from here we can put $s=0$ in
$\E_{\rm quant}^{\rm PV}$.}, and in the limit we are left with
\be\label{eq1PV}\E_{\rm quant}^{\rm PV}=
\frac{-a_2}{16\pi^2}\ln M+\frac{a_2}{8\pi^2}\ln m +\E_{\rm quant}^{\rm ren}(m).
\ee
Here we have taken into account that $\E_{\rm quant}^{\rm ren}(s,M)$
vanishes for $M\to\infty$.  Next, the contribution of $\ln M$ is
canceled from the corresponding contribution from the classical energy
by means of \Ref{g}. Finally, in \Ref{eq1PV} the limit $m\to0$ gives
a finite result. This follows because $\Eq$ as given by Eq.\Ref{eqd}
is finite for a massless theory.

For the total energy we obtain
\bea\label{eges1}\E&=&\E_{\rm class}+\E_{\rm quant}^{\rm PV} \nn \\
&=&-\frac{11N}{24\pi}\ \ln{\Lambda_{\rm QCD}} \
\int_0^\infty\frac{dr}{r} \ 
\mu'(r)^2 + \E_{\rm quant}^{\rm renPV}
\eea
with the notation
\be\label{eqfin}\E_{\rm quant}^{\rm renPV}=
\left(\frac{a_2}{16\pi^2}\ln m +\E_{\rm
  quant}^{\rm ren}(m)\right)_{_{\mid m=0}}
\ee
for the vacuum energy renormalized in the Pauli-Villars scheme. 

Eqs.\Ref{eges1} and \Ref{eqfin} are the final answer for the complete
energy of the vortex in QCD. It will be calculated for a number of
profile functions in the subsequent sections. Because the vortex
itself is unstable as its classical energy is positive a physical
interest is in configurations lowering the complete energy. To find
those a dimensional analysis can be made following
\cite{Diakonov:2002bx}. First we note that the profile $\mu(r)$ is a
dimensionless function of a variable with the dimension of a
length. Therefore it must contain an intrinsic scale $r_0$ so that in
fact we have to use
\be\label{mu0}\mu\to \mu\left(\frac{r}{r_0}\right).
\ee
Now we introduce the notation
\be\label{A}A= \pi\int_0^\infty\frac{dr}{r} \ \mu'(r)^2
\ee
to be used in Eq.\Ref{eges1}. Note that $A$ is dimensionless and a
  factor of $1/r_0^2$ had been taken out of the corresponding integral
  in Eq.\Ref{eges1}.  Next we note that $\E_{\rm quant}^{\rm renPV}$,
  Eq.\Ref{eqfin} has for a background profile as given by Eq.\Ref{mu0}
  the following dependence on $r_0$,
\be\label{2}\E_{\rm quant}^{\rm  renPV} =
-\frac{11N}{24\pi^2r_0^2}\ A \ \ln{r_0} +
\frac{1}{r_0^2} {\E_{\rm quant}^{\rm  renPV}}_{\mid r_0=1}.
\ee
This is for dimensional reasons.  We introduce by means of
\be\label{B}\ln B = 
-\frac{24\pi^2}{11N A}   \ {\E_{\rm quant}^{\rm  renPV}}_{\mid r_0=1}
\ee
the notation $B$, which allows to represent the complete energy in
the form
\be\label{eges2}\E
=-\frac{11N}{24\pi^2} \frac{A}{r_0^2} \ln\left(r_0\Lambda_{\rm QCD} B\right).
\ee
After this rewriting the dependence of the complete energy on the
scale $r_0$ is present in the shown places only. It is easy to find
the minimum as a function of $r_0$. After trivial calculation the
minimum appears at
\be\label{rmin}{r_0}_{\mid_{\rm min}}
=\frac{1}{\Lambda_{\rm QCD}}\frac{\sqrt{e}}{B},
\ee
where $e$ is the basis of the natural logarithms. The energy in the
minimum is
\be\label{egesmin}\E_{\mid_{\rm min}}=
-\Lambda_{\rm QCD}^2 \ \frac{11N}{24\pi^2} \ \frac{AB^2}{2e}.
\ee
This is the quantity to be compared for different profiles. Here the
sign of the vacuum energy enters as follows. 
Primarily, the sign of the complete energy in the minimum,
$\E_{\mid_{\rm min}}$, is always negative. This is a consequence of
the sign of $a_2$ resp. of the first coefficient of the beta
function. Secondly, the sign of $\E_{\rm quant}^{\rm renPV}$ in
Eq.\Ref{2} is dominated by $\ln r_0$ in the first contribution on the
rhs of Eq.\Ref{2} and depends in this way on the scale. The second
contribution, $ {\E_{\rm quant}^{\rm renPV}}_{\mid r_0=1}$, may be
positive or negative which will be a result of the calculations for a
specific profile $\mu(r)$. If the sign of $ {\E_{\rm quant}^{\rm
renPV}}_{\mid r_0=1}$ is positive (negative) we have $B<1$ ($B>1$) and
this contribution increases (decreases) the complete energy. At once,
$B$ influences the core radius in the minimum, ${r_0}_{\mid_{\rm
min}}$, making it larger (smaller) for $B>1$ ($B<1$).  We will see
below in all considered examples we have $B>1$. After this discussion
of the scale dependence in all following calculations (Sec. 3 and 4)
we put $r_0=1$.

\section{The finite   part of the vacuum energy}\label{Sec3}
The regularized vacuum energy is given by Eq.\Ref{eqr} and we need to
represent it in the form of Eq.\Ref{eqd}, i.e we need to separate the
pole part in $s$. Eq.\Ref{eqr} is well defined for $s\ge\frac32$ and
we need to perform the analytic continuation to $s=0$. In order to do
that we add and subtract a piece of the uniform asymptotic expansion
of the logarithm of the Jost function which we define by
\be\label{defas}
\ln f_\nu(ik)\sim \ln f^{\rm as}_\nu(ik)+O\left(\frac{1}{\nu^4}\right)
\ee
for $\nu$, $k\to\infty$ with $z\equiv\frac{k}{\nu}$ fixed. In this way
we represent
\be\label{Etot}\Eq=\Ef+\Eas+\Efn+\Easn
\ee
with
\be\label{Ef}
\Ef=\frac{-1}{8\pi }\sum_{c,\la}\sum_{\nu=-\infty \atop \nu\ne 0}^{\infty}  
\int_m^\infty dk \ 
\left(k^2-m^2\right)\frac{\pa}{\pa k} 
 \left( \ln f_\nu^{(c,\la)}(ik)-\ln f_\nu^{(c,\la){\rm as}}(ik)\right),
\ee
and
\be\label{Eas}
\Eas=\frac{C_s}{-8\pi }\sum_{c,\la}\sum_{\nu=-\infty \atop \nu\ne 0}^{\infty}  
\int_m^\infty dk \ 
\left(k^2-m^2\right)^{1-s}\frac{\pa}{\pa k} \ln f_\nu^{(c,\la){\rm as}}(ik),
\ee
where we in the Jost functions again indicated the dependence on color
and on polarisation.  In Eq.\Ref{Ef} we have put $s=0$ because there the
sum and the integral are convergent. This is ensured by the choice of
the asymptotic part of the Jost function. It can be checked that both,
the integral and the sum are made convergent by subtracting the
uniform asymptotic expansion. This fact is non trivial and doesn't
hold for any background potential. For instance, it can be shown that
for a $\mu(r)$ with a non vanishing first derivative in $r=0$ the
integral is not convergent.  But in that case the classical part of
the energy, Eq.\Ref{Eclass}, is infinite. This we do not consider.

In Eqs. \Ref{Ef} and \Ref{Eas} we left out the contribution from the
orbital momentum $\nu=0$ because it needs a separate treatment. Here
we define the asymptotic part of the Jost function as that for large
argument $k$,
\be\label{3}f_0(ik)\sim f_0^{\rm as}(ik)+O\left(\frac{1}{k^4}\right)
\ee
for $k\to\infty$. Accordingly we have
\be\label{Ef0}\Efn=\frac{-1}{8\pi }\sum_{c,\la}\int_m^\infty dk \ 
\left(k^2-m^2\right)\frac{\pa}{\pa k} 
\ln \left(f_0^{(c,\la)}(ik)-f_0^{(c,\la){\rm as}}(ik)\right),
\ee
and
\be\label{Eas0}\Easn=\frac{C_s}{-8\pi }\sum_{c,\la}
\int_m^\infty dk \ 
\left(k^2-m^2\right)^{1-s}\frac{\pa}{\pa k} \ln f_0^{(c,\la){\rm as}}(ik).
\ee
Again, in Eq.\Ref{Ef0} we could set $s=0$ because the integral over
$k$  is convergent.

The asymptotic expanson of the Jost function can be obtained in the
same way as in Ref.\cite{Bordag:1996fv} and in
Ref.\cite{Bordag:1998tg} from iterating the Lippmann-Schwinger
equation. While in \cite{Bordag:1996fv} the background was scalar only
two iterations were needed while in \cite{Bordag:1998tg} for a vector
background four had to be included. In the present case the
background is a vector, hence we need four iterations. In
\cite{Bordag:1998tg} the quantum field was a spinor one and here we
have a vector field. So we need to modify the formulas of
\cite{Bordag:1998tg} to some extend. The basic formula is the
Lippmann-Schwinger equation which appears if rewriting the wave
equation \Ref{radeq} as an integral equation (we drop the color and
polarization indices for a moment),
\be\label{LS}\phi(r)=\phi^0(r)+\frac{\pi}{2i}\int\limits_0^r\frac{dr'}{r'}
\left(  J_\nu(kr) H^{(1)}_\nu(kr')- 
        J_\nu(kr')H^{(1)}_\nu(kr ) \right)P(r') \phi(r')
\ee
with the 'perturbation operator'
\be\label{P}P(r)=-2\al_c\mu(r)+\al_c^2\mu(r)^2-2\al_c\beta_\la r\mu'(r)
\ee
and the choice  $ \phi^0(r)=J_\nu(kr)$ for 'free' solution which just
makes $\phi(r)$ defined by Eq.\Ref{LS} the regular solution. From
considering this equation for $r\to\infty$ and comparing with
Eq.\Ref{defJ} the following representation for the Jost function can
be obtained,
\be\label{LSJ}f_\nu(k)=1+\frac{\pi}{2i}\int\limits_0^\infty\frac{dr}{r}
H^{(1)}_\nu(kr ) P(r) \phi(r).
\ee

The next steps go exactly as in Ref.\cite{Bordag:1998tg}. The
logarithm of the Jost function which is the quantity we are interested
in can be represented as given by Eqs.(32)-(35) in
\cite{Bordag:1998tg} with obvious substitutions. In these formulas one
has to turn to imaginary $k$. After that   the modified Bessel
functions entering have to be substituted by their well known uniform
asymptotic expansion and the integrations have to be performed by the
saddle point method keeping all the contributions up to the necessary
order. This is a quite lengthy calculation which is best performed by
a computer algebra package to avoid errors. The result is a
representation of the asymptotic expansion of the logarithm of the
Jost function of the form   
\be\label{Jexp}\ln f_\nu^{\rm as}(ik)=
\sum_{n=1}^3\sum_{j=n}^{3n} \int\limits_0^\infty\frac{dr}{r}
 X_{nj}\frac{t^j}{\nu^n}
\ee
with
\be\label{t}t=\frac{1}{\sqrt{1+(z r)^2}}
\ee
($z=k/\nu$). The coefficients $X_{nj}$ are local functions of $\mu(r)$
and its derivatives. They are listed in the Appendix. It turns out
that this is not the most convenient form. It is possible to integrate
in Eq.\Ref{Jexp} by parts so that all contributions become proportional at
last to one derivative of $\mu(r)$. In particular this speeds up the
convergence of the integrals at $r\to\infty$ which later on must be
taken numerically. After integrating by parts the formula looks
essentially the same except that the coefficients are different. We
denote them by $Z_{nj}$ so that we obtain $\ln f_\nu^{\rm as}(ik)$ in
the form 
\be\label{asZ}\ln f_\nu^{\rm as}(ik)=
\sum_{n=1}^3\sum_{j=n}^{3n} \int\limits_0^\infty\frac{dr}{r}
 Z_{nj}\frac{t^j}{\nu^n},
\ee
where the new coefficients are listed in the Appendix too. 

For zero orbital momentum, $\nu=0$, we need the asymptotic expansion
of the Jost function for large argument. It can be obtained from
Eqs.(32) and (33) in Ref.\cite{Bordag:1998tg} by inserting there the
asymptotic expansion of the Bessel functions for large argument. This
is much easier  than in the case of the uniform asymptotic
expansion. This allows us to present these formulas here in order to
illustrate the method. From the mentioned equations in
\cite{Bordag:1998tg} we have
\be\label{lnf1}\ln f^{(1)}=
\int\limits_0^\infty\frac{dr}{r}I_\nu(kr)K_\nu(kr)P(r)
\ee
and
\be\label{lnf2}\ln f^{(2)}=
\int\limits_0^\infty\frac{dr}{r}K_\nu^2(kr)P(r)
\int\limits_0^\infty\frac{dr'}{r'}I_\nu^2(kr')P(r') .  
\ee
In Eq.\Ref{lnf1} we use the expansion
\be\label{4}I_\nu(kr)K_\nu(kr)=\frac{1}{2kr}+\frac{1}{16(kr)^3}+\dots \,.
\ee
In Eq.\Ref{lnf2} it is sufficient to use the leading order from the
expansion of the Bessel functions,
$K_\nu^2(kr)=\exp(-2kr)/(2kr)+\dots$ and
$I_\nu^2(kr)=\exp(2kr)/(2kr)+\dots$ and to take the integral over $r'$
by the saddle point method to leading order. The result is
\be\label{lnf2a}\ln f^{(2)}=-\frac{1}{8k^3}
\int\limits_0^\infty\frac{dr}{r} P(r)^2.
\ee
Taking both contributions together we define
\be\label{asP}\ln f^{\rm as}_0(ik)=
\int\limits_0^\infty\frac{dr}{r} \left\{
\frac{P(r)}{2kr}+\frac{1}{\left(k^2+1\right)^{3/2}}
\left(\frac{P(r)}{16r^3}- \frac{P^2(r)}{8r^3}\right) \right\}.
\ee
Here we substituted $1/k^3$ by $1/(k^2+1)^{3/2}$. This does not change
the asymptotics to the given order but avoids problems with the
integration over $k$ at the origin in Eq.\Ref{Ef0} when we put $m=0$
later on.

Now, in order to calculate the finite parts of the vacuum energy we
need a method to calculate the Jost function numerically.  Here we
follow the ideas of Ref.\cite{Bordag:1999sf} with necessary
modifications. The Jost function $f_\nu(k)$ is defined by
Eq.\Ref{defJ}. We start from defining a new function $f_\nu(k,r)$ by
means of
\be\label{J1}\phi(r) =
\frac12\left(  f_\nu(k,r)\  H^{(2)}_{\nu-\al_c}(kr)+ 
\overline{f}_{\nu}(k,r) \ H^{(1)}_{\nu-\al_c}(kr)        \right).
\ee
From the definition, Eq.\Ref{defJ}, it is clear that
\be\label{5} f_\nu(k,r) \ 
{\raisebox{-7pt}{$\sim$}\atop  {\mbox{$\scriptstyle   r\to\infty$}}} \
f_\nu(k).
\ee
The speed of convergence depends on how fast the potential $\mu(r)$
approaches its value at infinity. It can be shown that the derivative
of $ f_\nu(k,r)$ with respect to $r$ is proportional to $\mu(r)-1$.

For a numerical calculation of the wave function $\phi(r)$ we can use
the radial equation, Eq.\Ref{radeq} and solve it with some Runge-Kutta
procedure. For this end we need initial conditions at $r=0$.  Here it
is necessary to introduce a new function by means of
\be\label{6}\phi(r)=
\left(\frac{kr}{2}\right)^\nu \ \frac{\varphi(r)}{\Gamma(\nu+1)}.
\ee
Given $\phi(r)$ is the regular solution it behaves for $r\to0$ like the
Bessel function $J_\nu(kr)$ and the introduced function $\varphi(r)$ is
finite at the origin,
\be\label{7}\varphi(0)=1.
\ee
As the second initial value we need the derivative of $\varphi(r)$ in
$r=0$. Because $\mu(r)$ vanishes there together with its first
derivative,
\be\label{8}\varphi'(0)=0
\ee
holds. 

In the numerical calculations we used Mathematica and the build in
function \verb!NDSolve!.  Here it is useful to turn to a system of
first order equation by means of
\beao\label{dgl}\varphi'(r)&=&\psi(r), \\
\psi'(r)&=&-\frac{2\nu+1}{r}\psi(r)
+\left(-\frac{2\al_c\nu\mu(r)}{r}+\frac{\al_c^2\mu(r)^2}{r^2}
-\frac{2\al_c\beta_\la\mu'(r)}{r}-k^2
\right)\varphi(r).
\eeao
Also it is necessary to move the starting point of the procedure away
from $r=0$ because otherwise \verb!NDSolve! tries to divide
$\varphi'(0)$ by $r=0$. The initial data shifted by a small amount
$\Delta$ can be obtained from a decomposition of the equations in
powers of $r$ and in the lowest nontrivial order we get
\bea\label{indat}\varphi(\Delta)&=&1\nn, \\
\psi(\Delta)&=&\Delta \ \mu_2 
\eea
with $\mu_2=\mu''(r)_{\mid{r=0}}$. Now it remains to express the Jost
function, $f_\nu(k,r)$, from Eq.\Ref{J1} in terms of the functions
$\varphi(r)$ and $\psi(r)$. In addition we need the Jost function for
imaginary argument. So we have to substitute $k\to ik$ in the above
equations.  An easy calculation gives
\be\label{fnu}f_\nu(ik,r)=\frac{\left({kr/2}\right)^\nu} {\Gamma(\nu+1)}
\left(
r \psi(r) K_{\nu-\al}(kr)+\varphi(r)
\left(k K_{\nu+1-\al}(kr)+\frac{\al}{r}K_{\nu-\al}(kr)\right)  \right),
\ee
where $K_{\nu-\al}(kr)$ are modified Bessel functions.  In the process
of solving the equations using \verb!NDSolve! the true Jost function
is approached quite fast. In all considered examples it was sufficient
to take $r=10$.

By means of Eqs.\Ref{Ef} and \Ref{asZ} we have all we need for a
numerical calculation of $\Ef$. In the actual calculations we used
Eq.\Ref{Ef} in a modified form. We integrated by parts and change the
integration variable from $k$ to $z=k/\nu$. Finally, we put $m=0$
because here this auxiliary mass is no longer needed. In this way the
formula reads
\be\label{Ef1}\Ef=\frac{1}{2\pi}
\sum_{c,\la}\sum_{\nu=-\infty \atop \nu\ne 0}^{\infty}  \nu^2
\int_0^\infty dz \ z \
\left(\ln  f_\nu^{(c,\la)}(i\nu z)-\ln f_\nu^{(c,\la){\rm as}}(i\nu z)\right).
\ee
Here a number of comments are in order. First, in Eq.\Ref{fnu} is
written for positive orbital momenta. This is sufficient because the
negative ones can be obtained by changing the signs of $\al_c$ and
$\beta_\la$ as can be seen from Eq.\Ref{radeq}. Second, there are
cancellations between contributions with different signs of the
orbital momenta. For instance, there are even powers in $1/\nu$ in the
uniform asymptotic expansion. They cancel\footnote{This is taken into
  account already in the coefficients $X_{nj}$ and $Z_{nj}$ listed in
  the Appendix.}. Also, contributions odd in $\beta_\la$ cancel in the
sum over the polarizations. For large $\nu$, because of the definition
Eq.\Ref{defas} the difference of the two logarithms in Eq.\Ref{Ef1}
decreases as $1/\nu^4$. In fact, under the sum over  color and
polarizations it decreases as $1/\nu^5$. As a function of $z$ it
decreases as $1/z^4$. The integral over $z$ decreases as $1/\nu^5$. So
the convergence is quite well. For instance, it is sufficient to take
a few orbital momenta to achieve a precision of several digits. 

\begin{figure}[t] 
\epsfxsize=11cm
\epsffile{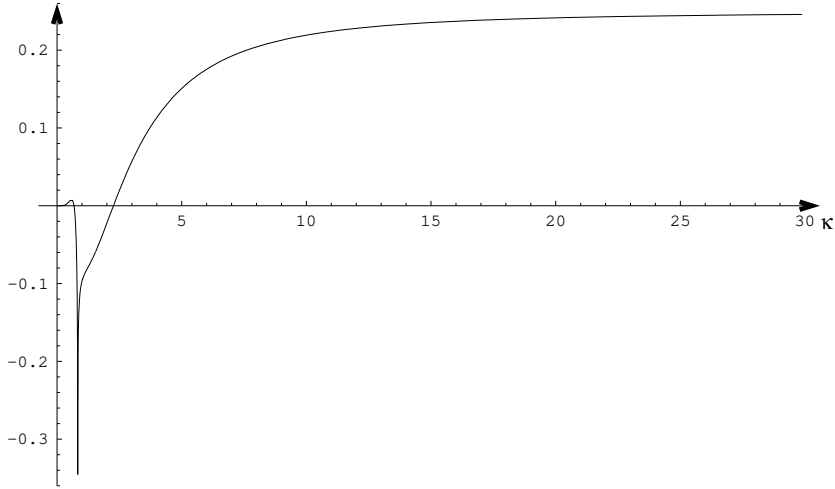}
\caption{The s-wave contribution, $\Efn(k)$, multiplied by $k^4$.}
\end{figure}
\begin{figure}[h] 
\epsfxsize=11cm
\epsffile{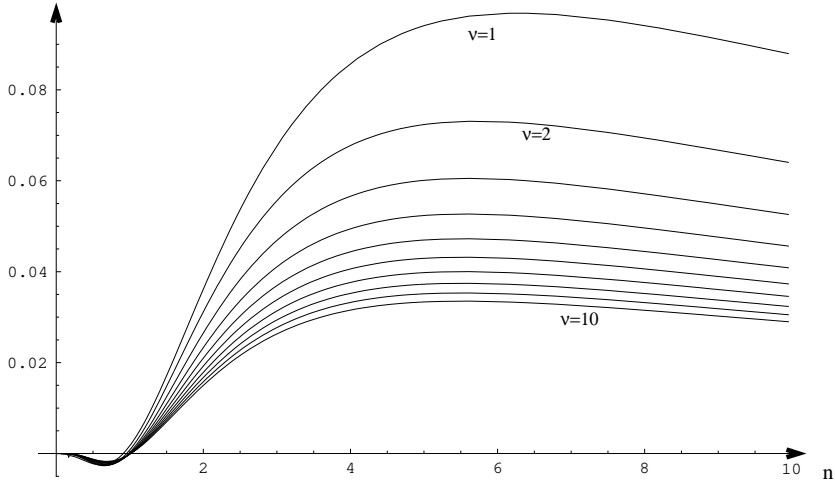}
\caption{The functions   $\Ef(\nu,z)$   multiplied by $z^3\nu^{4.5}$
for the first ten orbital momenta.}\label{fig1}
\end{figure} %
\begin{figure}[h] 
\epsfxsize=11cm
\epsffile{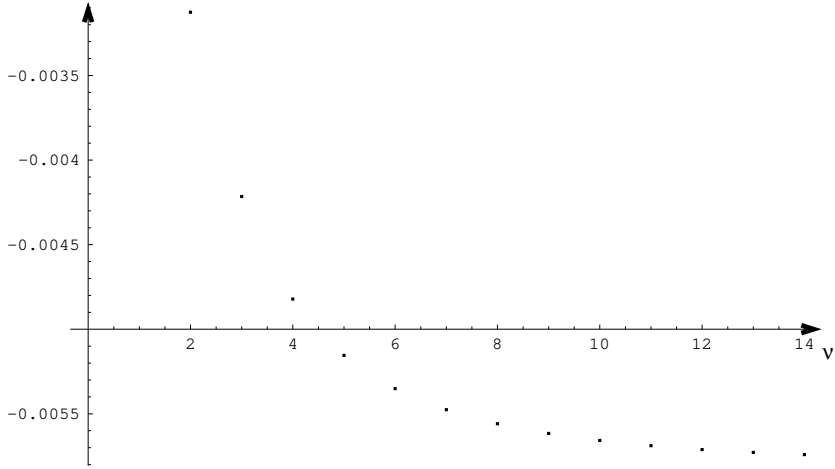}
\caption{The contributions $\Ef(\nu)$ to the vacuum energy multiplied
 by $\nu^5$ from the higher orbital momenta}
\end{figure}
 
In order to represent the results we introduce the notations
\be\label{9}\Ef(\nu,z)=\frac{z}{\pi}   \sum_{c,\la}
\left(\ln  f_\nu^{(c,\la)}(i\nu z)-\ln f_\nu^{(c,\la){\rm as}}(i\nu z)\right)
\ee
where the factor of 2 resulting from the two signs of the orbital
momentum was taken into account and
\be\label{Efrep1}\Ef(\nu)=  \int\limits_o^\infty dz \ z\   \Ef(\nu,z)
\ee
where the factor of 2 results from the two sign of the orbital
momentum so that
\be\label{Efrep2}\Ef=\sum_{\nu=1}^\infty \nu^2 \Ef(\nu)
\ee
holds.

In a similar way the contribution of the s-wave ($\nu=0$) can be
treated. Actually, it is much easier because there is only one
integration and no sum.  We put $m=0$ in Eq.\Ref{Ef0}, integrate by
parts and write it in the form
\be\label{Ef0rep1}\Efn=\int\limits_0^\infty dk\ k\ \Efn(k)
\ee
with
\be\label{Ef0rep2}\Efn(k)=\frac{k}{4\pi}\ \sum_{c,\la} \ln 
\left(f_0^{(c,\la)}(ik)-f_0^{(c,\la){\rm as}}(ik)\right).
\ee
As an example let us consider the profile
\be\label{muex1}\mu(r)=\tanh^2(r).
\ee
It vanishes for $r\to0$ together with its first derivative and it
approaches its value at infinity exponentially fast.

Fig. 1 shows the function $\Efn(k)$ multiplied by $k^4$. It can be
seen that its decrease is confirmed by the numerical results.

In Fig. 2 the functions $\Ef(\nu,z)$ multiplied by $z^3\nu^{4.5}$ are
plotted showing that the expected behavior for large $z$ is confirmed
by the numerical evaluation. The factor $z^3$ is taken to demonstrate
the fast asymptotic decrease and the factor $\nu^{4.5}$ in order to
have the curves for all orbital momenta ($\nu=1,\dots,10$) to fit into
one figure. 

In Fig. 3 the contributions $\Ef(\nu)$ from the higher orbital momenta
($\nu=2,\dots,10$) to the vacuum energy multiplied by $\nu^5$ are
shown. The corresponding sum in Eq.\Ref{Ef0rep1} is fast converging and
in fact it is sufficient to take some few first orbital momenta
only. As an illustration we show the corresponding numbers,
\be\label{Ex1}\begin{tabular}{rcccc}
\multicolumn{5}{l}{$\Efn$=-0.0482214,}\\
$\nu$= &1&2&3&4,\\
$\nu^2\Ef(\nu)$= &-0.00587755& -0.00039075& -0.000156122& -0.0000753351.
\end{tabular}
\ee

It should be remarked that the observed decrease of the subtracted Jost
functions is a strong check confirming that the formulas for both, the
Jost function itself and for its asymptotic expansions, are
correct. This is because these formulas are obtained in completely
different ways. The first one is a numerical solution of the
differential equation and the second one results form the asymptotic
expansion which we obtained from the Lippmann-Schwinger equation. 

Further, it should be noticed that in each case the decrease is by one
power better than it follows from what we subtracted. This is because
each second power disappears under the sum of color and polarization.

As already mentioned, it turns out that for $\nu=0$ and for $\nu=1$
there are bound states. They manifest themselves as poles of the
logarithm of the Jost function and, as a consequence, of the functions
$\Efn(k)$ and $\Ef(\nu,k)$.  These functions are shown in Fig. 4.

\begin{figure}[t]  
\epsfxsize=11cm
\epsffile{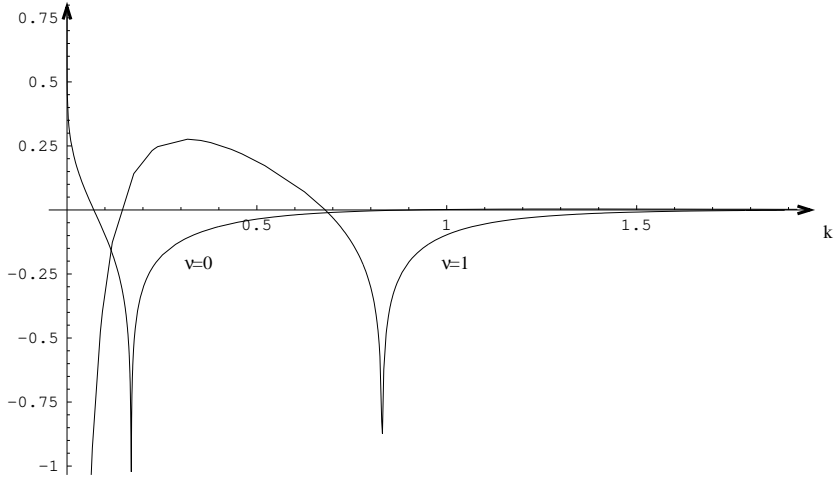}
\caption{The functions $\Efn(k)$ and    $\Ef(1,k)$   for small $k$. The
  poles indicate the positions of the bound states.} 
\end{figure}

\section{The  asymptotic part of the vacuum energy}\label{Sec4}
Next we have to calculate the asymptotic contributions to the vacuum
energy. Actually, the problem is to remove the auxiliary mass $m$ and
perform the analytic continuation to $s=0$.  We start with $\Easn$,
Eq.\Ref{Eas0} and insert there the asymptotic expansion \Ref{asP}. For
$s>\frac12$ we perform the limit $m\to0$. The contribution from $\ln
f^{\rm as}_0(ik)$ which is proportional to $1/k$ vanishes. After that
we can put $s=0$ and are left with the finite integral
\[ \int_0^\infty dk\ k^2 \frac{\pa}{\pa k}\frac{1}{(k^2+1)^{3/2}}=-2
\]
and obtain
\be\label{10}\Easn=\frac{1}{-8\pi }\sum_{c,\la}
\int\limits_0^\infty\frac{dr}{r}
\left(\frac{P(r)}{16r^3}- \frac{P^2(r)}{8r^3}\right),
\ee
where $P(r)$ is given by Eq.\Ref{P}.
The sum over color and polarization can be carried out using \Ref{alc}
and \Ref{betla},
\be\label{sumcl}\begin{array}{rclrclrcl}
\sum\limits_{c,\la} 1&=& 2N , \qquad 
\sum\limits_{c,\la} \al_c^2&=& 2N,  \qquad
\sum\limits_{c,\la} \al_c^4&=& 4 \ \ \mbox{(SU(2))}, \\[14pt]
\sum\limits_{c,\la} \al_c^4&=& 9/2 \ \ \mbox{(SU(3))}, && 
\sum\limits_{c,\la} \al_c^2\left(\frac{1}{12}-\beta_\la^2\right)
&=&  \frac{-11N}{6}.
\end{array}\ee
We obtain
\be\label{Easnullp}\Easn=\frac{N}{32\pi}\int\limits_0^\infty\frac{dr}{r}
\left(\mu^2(r)-\frac{N}{4}\mu^4(r)-8 r^2 (\mu'(r))^2 \right)
\ee
for the s-wave contribution to the asymptotic part of the vacuum
energy. 

The contributions from the higher orbital momenta require more
work. We start from rewriting $\Eas$, Eq.\Ref{Eas}, using the
Abel-Plana formula (see Appendix, Eq.\Ref{APla}) in the form
\be\label{Easz}\Eas=\sum_{c,\la}
\left\{ \E^{\rm as,\ a} -\frac12 \E^{\rm as,\ 0} + \E^{\rm as,\ b} \right\}
\ee
with
\bea\label{Easza} \E^{\rm as,\ a}&=&  \frac{C_s}{-4\pi }
\int_0^\infty d\nu\ \int_m^\infty dk \ 
\left(k^2-m^2\right)^{1-s}\frac{\pa}{\pa k} \ln f_\nu^{(c,\la){\rm as}}(ik),
\\
\label{Easz0}\E^{\rm as,\ 0}&=&  \frac{C_s}{-4\pi }
\lim_{\nu\to0}     \int_m^\infty dk \ 
\left(k^2-m^2\right)^{1-s}\frac{\pa}{\pa k} \ln f_\nu^{(c,\la){\rm as}}(ik),
\\
\label{Easzb}\E^{\rm as,\ b} &=&   \frac{C_s}{-4\pi }
\int\limits_{0}^{\infty}{d \nu\over 1-e^{2\pi\nu}}
\int_m^\infty dk \ 
\left(k^2-m^2\right)^{1-s}\frac{\pa}{\pa k} \\ && \hspace{6cm} \nn \times
\frac{1}{i}\left( \ln f_{(i\nu)}^{(c,\la){\rm as}}(ik) -\mbox{(c.c.)} \right),
\eea
where (c.c.) denotes complex conjugate. It should be noticed that
$f_\nu^{(c,\la){\rm as}}(ik)$ is real so that (c.c.) changes in fact
the sign of $i\nu$ only. In Eq.\Ref{Easz} a factor of 2 from the
negative orbital momenta was taken into account. The contribution from
$\nu=0$ deserves a remark. In fact, Eq.\Ref{APla} is valid for a
finite $f(0)$ only. This is ensured in our case by the auxiliary mass
$m$. So we have first to perform the limes $\nu\to0$. After that the
limit $m\to0$ produces from $\E^{\rm as,\ 0}$ a divergent contribution
proportional to $1/m$ which is, however, canceled by a corresponding
contribution from $\E^{\rm as,\ b}$. 

Now we use the uniform asymptotic expansion of the logarithm of the
Jost function as given by Eq.\Ref{asZ}. So we insert it into $\E^{\rm
as,\ a}$ and interchange the orders of integration. For each
contribution of the sum over $n$ and $j$ we use formula \Ref{C2}
given in the Appendix and obtain the representation
\be\label{11}\E^{\rm as,\ a}  =
\frac{1+s(2\ln2-1)}{8\pi } m^{-2s} \Gamma(2-s)\Gamma(s)
\int\limits_0^\infty\frac{dr}{r^3}
\sum_{c,\la}\sum_{j=3}^{7}\frac{Z_{3j}}{\frac{j}{2}-1},
\ee
where contributions proportional to powers of $m$ are dropped. 
From Eqs. \Ref{Znj} and \Ref{sumcl} we obtain
\be\label{12}\sum_{c,\la}
\sum_{j=3}^{7}\frac{Z_{3j}}{\frac{j}{2}-1}
=\frac{-11N}{6}\left(r\mu'(r)\right)^2.
\ee
In the limit $s\to0$, $\E^{\rm as,\ a}$ takes the form
\be\label{Easa}\E^{\rm as,\ a}=\frac{-11N}{48\pi^2}\ A\
\left(\frac{1}{s}+2\ln2-1-\gamma-2\ln m+O(s)\right)  
\ee
with $A$ given by Eq.\Ref{A}. 

Next contribution to be considered is $\E^{\rm as,\ 0}$,
Eq.\Ref{Easz0}. We insert $\ln f_\nu^{\rm as}(ik)$ from Eq.\Ref{asZ}
and use formula \Ref{C3} from the Appendix,
\bea\label{13}\E^{\rm as,\ 0}&=&
\frac{C_s)}{4\pi } 
\lim_{\nu\to0} 
\int\limits_0^\infty\frac{dr}{r}
\sum_{c,\la}\sum_{n=1}^3\sum_{j=3n}^{3n} Z_{nj} \nn \\ && \times
 {\Gamma(2-s)\Gamma(s+\frac{j}{2}-1)\over (mr)^j\ \Gamma(\frac{j}{2})}
{m^{2-2s} \left(  {\nu }\right)^{j-n}\over 
\left(1+\left({\nu\over mr}\right)^{2}\right)^{s+\frac{j}{2}-1}}.
\eea
Here in the limes $\nu\to0$ only contributions with $j=\nu$ remain
(note $j\ge n$ in Eq.\Ref{Jexp}). After that we consider the limits
$m\to0$ and   $s=0$. We obtain
\be\label{Easnull}\E^{\rm as,\ 0} =
\frac{1}{2\pi m}  \int\limits_0^\infty\frac{dr}{r^4}
\sum_{c,\la}  Z_{33}   +O(m).
\ee
The contribution from $n=2$ is odd under $\beta_\la\to-\beta_\la$ and
does not contribute.

It remains to consider $\E^{\rm as,\ b}$, Eq.\Ref{Easzb}. Again,
inserting $\ln f_\nu^{\rm as}(ik)$ from Eq.\Ref{asZ} and using formula
\Ref{C3} from the Appendix,
\bea\label{Easb}\E^{\rm as,\ b}    &=&
\frac{C_s)}{4\pi } 
\int_m^\infty dk\ (k^2-m^2)^{1-s} \frac{\pa}{\pa k}
\int\limits_0^\infty\frac{dr}{r}  \sum_{n=1}^3\sum_{j=n}^{3n}  Z_{nj} 
\nn \\ && \times   m^{2-2s} \
{\Gamma(2-s)\Gamma(s+\frac{j}{2}-1)\over (mr)^j\ \Gamma(\frac{j}{2})}\
\Sigma_{nj}(mr)
\eea
with the notation 
\be\label{14}\Sigma_{nj}(x)=
\int\limits_{0}^{\infty}{d \nu\over 1-e^{2\pi\nu}}
\frac{1}{i}  \left[  
\frac{(i\nu)^{j-n}}{\left(\left(\frac{i\nu}{x}\right)^2+1\right)^{s+j/2-1}}
-\mbox{(c.c.)}\right]
\ee
holds. In $\Sigma_{nj}(x)$ there is in fact no contribution from
$\nu\le x$ because $j-n$ is even. It can be rewritten in the form
\be\label{15}\Sigma_{nj}(x)=
\int_x^\infty{d \nu\over 1-e^{2\pi\nu}} \ \ 
\frac{(-1)^{(n-1)/2}\cos\pi s \ 
\nu^{j-n}}{\left(\left(\frac{\nu}{x}\right)^2-1\right)^{s+j/2-1}}.
\ee
We have to perform the analytic continuation to $s=0$. For $n=j=1$ we
simply can put $s=0$ and obtain for small $x$ (this is because we need
$m\to0$) 
\be\label{sig11}\Sigma_{11}(x)=\frac{-1}{12x}+O(x).
\ee
For $n=j=3$ we also can put $s=0$ and obtain
\be\label{16}\Sigma_{33}(x)=
x\int_x^\infty{d \nu\over 1-e^{2\pi\nu}} \frac{1}{\sqrt{\nu^2-x^2}}.
\ee
This integral is convergent. But for $x\to0$ this form is not
convenient. So we integrate two times by parts. We obtain 
first
\be\label{17}\Sigma_{33}(x)=
\frac{-\pi x}{1-e^{2\pi x}}
-2\int_x^\infty d\nu\ \arctan\frac{x}{\sqrt{\nu^2-x^2}} 
\left(\frac{\nu}{1-e^{2\pi\nu}}\right)'
\ee
and subsequently
\bea\label{18}&&\Sigma_{33}(x)=
\frac{-\pi x}{1-e^{2\pi x}}
+x\left(\pi+2\ln2x^2\right)\left(\frac{x}{1-e^{2\pi x}}\right)'   \\
\nn && 
+2 \int_x^\infty d\nu\ 
\left[ \nu \arctan\frac{x}{\sqrt{\nu^2-x^2}} + x\left( \ln 2x +\ln
  \left(\nu+\sqrt{\nu^2-x^2}\right)\right)\right]
\left(\frac{\nu}{1-e^{2\pi\nu}}\right)''.
\eea
Here we can tend $x$ to zero under the sign of the integral and arrive at
\be\label{sig33}\Sigma_{33}(x)=\frac12+x\left(\ln x + \Sigma_3  \right) +O(x^2)
\ee
with
\be\label{19}\Sigma_3 \equiv 
-1+
2 \int_0^\infty d\nu\ \ln (2\nu)\
\left(\frac{\nu}{1-e^{2\pi\nu}}\right)'' \ =\ -1.36437.
\ee
For $n=3$, $j=5$ we proceed in a similar way. In order to perform the
analytic continuation to $s=0$ we integrate by parts and obtain
\be\label{20}\Sigma_{35}(x)=-2x^3  \int_x^\infty
\frac{d\nu}{\sqrt{\nu^2-x^2}}
\left(\frac{\nu}{1-e^{2\pi\nu}}\right)'.
\ee
For $x\to0$ we have to integrate by parts once more and get
\be\label{21}\Sigma_{35}(x)= x^3\left(
\ln x+ \Sigma_5 \right) +O(x^4)
\ee
with
\be\Sigma_5\equiv 2
 \int_0^\infty d\nu\ \ln (2\nu)\
\left(\frac{\nu}{1-e^{2\pi\nu}}\right)'' \ =\ -0.364378  .
\ee
Finally we have $n=3$, $j=7$. For $s=0$ we integrate two times by
parts and obtain
\be\label{22}\Sigma_{37}(x)=-\frac23 x^5  \int_x^\infty
\frac{d\nu}{\sqrt{\nu^2-x^2}}
\left(\frac{1}{\nu}\left(\frac{\nu^3}{1-e^{2\pi\nu}}\right)'\right)'.
\ee
For $x\to0$ we have to integrate by parts once more and get
\be\label{23}\Sigma_{37}(x)= x^5\left(
\ln x+\Sigma_7\right)+O(x^6)
\ee
with
\be
\Sigma_7=
\frac23 \int_0^\infty d\nu\ \ln (2\nu)\
\left(\frac{1}{\nu}\left(\frac{\nu^3}{1-e^{2\pi\nu}}\right)'\right)''
\ =\  -0.121459 .
\ee

Having now the appropriate representations for $\Sigma_{nj}$ we can
insert them into $\E^{\rm as,\ b} $, Eq.\Ref{Easb}. We start from
considering $n=j=1$. 
Inserting \Ref{sig11} we obtain 
\be\label{Easb11}\E^{\rm as,\ b(11)}=
\frac{N}{48\pi} \int_0^\infty\frac{dr}{r^2} \  \mu(r)\mu'(r).
\ee
Next we look for the first contribution in the rhs. of
Eq.\Ref{sig33}. It gives a contribution to $\E^{\rm as,\ b} $,
Eq.\Ref{Easb} which is proportional to $1/m$ and it cancels just the
contribution of $\E^{\rm as,\ 0}$, Eq.\Ref{Easnull}. Next we consider
contributions from $\Sigma_{3j}$ containing the logarithm
$\ln(rm)$. What is proportional to $\ln m$ can be seen to cancel
exactly the corresponding contribution from $\E^{\rm as,\ a}$,
Eq.\Ref{Easa}. The remaining part is
\be\label{Easbln}\E^{\rm as,\ b({\rm ln})}=
\frac{-11N}{24\pi} \int_0^\infty\frac{dr}{r} \ \ln r\ \mu'(r)^2.
\ee
Finally we collect the contributions from $\Sigma_j$ and denote them
$\E^{\rm as,\ b(3j)}$ ($j=3,5,7$). They read explicitely
\be\label{24}\E^{\rm as,\ b(3j)}=
\frac{1}{4\pi}    \int_0^\infty\frac{dr}{r^3} \sum_{c,\la} \sum_j
\frac{Z_{3j}}{j/2-1} \Sigma_j
\ee
with $Z_{3j}$ given by Eqs.\Ref{Znj}. Performing the sum over color
and polarization we obtain
\bea\label{Easb3j}\E^{\rm as,\ b(33)}&=&
\frac{N\Sigma_3}{24\pi}
 \int_0^\infty\frac{dr}{r^2}
\left(-11\mu(r)\mu'(r)+6r\mu(r)\mu''(r)-6r(\mu'(r))^2 \right.\nn \\ && \left.
\hspace{6cm}  -\frac{N}{2} \mu^3(r)\mu'(r)\right), \nn\\
\E^{\rm as,\ b(35)}&=&
\frac{N\Sigma_5}{24\pi}
 \int_0^\infty\frac{dr}{r^2}
\left(26\mu(r)\mu'(r)-6r\mu(r)\mu''(r)-5r(\mu'(r))^2 \right.\nn \\ && \left.
\hspace{6cm}  +N \mu^3(r)\mu'(r)\right), \nn\\
\E^{\rm as,\ b(37)}&=&
 \frac{-5N\Sigma_7}{16\pi}
 \int_0^\infty\frac{dr}{r^2} \  \mu(r)\mu'(r) .
\eea

In this way the complete asymptotic part $\Eas$ from the higher
orbital momenta, Eq.\Ref{Easz}, of the
vacuum energy reads
\be\label{Easf}\Eas=
\E^{\rm as,\ b(11)}
+\E^{\rm as,\ b({\rm ln})}  
+\E^{\rm as,\ b(33)}
+\E^{\rm as,\ b(35)}
+\E^{\rm as,\ b(37)},
\ee
where the individual parts are given by Eqs. \Ref{Easb11},
\Ref{Easbln} and \Ref{Easb3j}. 

\begin{table}[h]
\begin{tabular}{|cll|}\hline&&\\[-9pt]
$\mu(r)$ & $\nu=0$ &$\nu=1$\\[7pt]\hline
$\tanh^2r$  & $\kappa_0=$ 0.82862  &$\kappa_1=$ 0.16771\\            
$\tanh^4r$ & $\kappa_0=$ 0.58388    &$\kappa_1=$ 0.16379 \\
$\tanh^8r$ &  $\kappa_0=$  0.46125    &$\kappa_1=$ 0.15486 \\
$r^6/(1+r^6)$ &$\kappa_0=$ 0.742808    &$\kappa_1=$ 0.28539 \\\hline
\end{tabular}\caption{The bound state levels $\kappa$ in Eq.\Ref{eqim}.}
\end{table}

\section{Numerical Results}\label{Sec5}
 
%
\begin{table}[h]\label{tab1}
\begin{tabular}{|ccccccc|}\hline&&&&&&\\[-9pt]
N& $ \E^{\rm as,\ b(11)} $ & $\E^{\rm as,\ b({\rm
      ln})}  $ & $\E^{\rm as,\ b(33)} $ & $\E^{\rm as,\ b(35)}$ &
      $\E^{\rm as,\ b(37)}$ & $\Eas$ \\ \hline  &&&&&&\\[-5pt]
\multicolumn{7}{|c|}{$\mu(r)=\tanh^2r$} \\
2&0.00823767& 0.16906& 0.507441& -0.138928& 0.0150082& 0.208057\\
3& 0.0123565& 0.25359& 0.753783& -0.204451& 0.0225122& 0.311155\\
\hline&&&&&&\\[-5pt]
\multicolumn{7}{|c|}{$\mu(r)=\tanh^4r$} \\
2& 0.00360497& 0.00354893& 0.272918& -0.0634807& 0.00656787& 0.138266 \\
3& 0.00540745& 0.00532339& 0.405012& -0.0928895& 0.00985181& 0.2064
\\\hline&&&&&&\\[-5pt]
\multicolumn{7}{|c|}{$\mu(r)=\tanh^8r$} \\
2&0.00213291& -0.0326859& 0.190348& -0.0386886& 0.00388594& 0.0825228\\
3&0.00319937& -0.0490288& 0.282582& -0.0564628& 0.00582891& 0.122963
\\\hline&&&&&&\\[-5pt]
\multicolumn{7}{|c|}{$\mu(r)=r^6/(1+r^6)$} \\
2&0.00534584& 0.0228004& 0.561348& -0.0998484& 0.00973955& 0.310584\\
3&0.00801875& 0.0342006& 0.833918& -0.145444& 0.0146093& 0.464541\\\hline
\end{tabular}\caption{The constituents of the asymptotic part $\Eas$ of the
  vacuum energy according to Eq.\Ref{Easf}.}
\end{table}

The actual calculation have been performed for four profiles $\mu(r)$,
see for example Table 1, using Mathematica. For the solution of the
differential equation \Ref{dgl} the build in function \verb!NDSolve!
had been used. The working precision was taken to be 25 digits and the
pure computation time was about 40 hours on a standard PC. The good
convergence of the integrals and of the sums was already discussed in
Sect. 3. The numerical integrations had been done with the build in
function which calculated about 200 points for each integral. In this
way the numerical effort is not very demanding.

   In the examples considered here there are bound states in the
spectrum of the fluctuations. They result in an imaginary part in the
vacuum energy, see Eq.\Ref{eqim}.  In the subdivision of the vacuum
energy according to Eq.\Ref{Etot} the imaginary part resides in the
finite parts, $\Efn$ and $\Ef$.  Thereby the first bound state appears
in the s-wave contribution $\Efn$, the second in the p-wave ($\nu=1$).
The binding energies are shown in Table 1.  These bound states do not
depend on the gauge group $SU(2)$ or $SU(3)$ because they come from
the color eigenvalues $\al_c=\pm1$ in Eq.\Ref{alc} which are common to
both groups.

In Table 2 the numerical results for the asymptotic parts of the
vacuum energy $\Eas$, Eq.\Ref{Eas} are shown. It is seen that $\E^{\rm
as,\ b(33)} $ is the numerically dominating contribution. Within it
the dominating part is $(\mu'(r))^2$ over weighting that of
$(\mu(r))^2$. Therefore there seems to be no chance to change the sign
of $\E^{\rm as,\ b(33)} $ taking another profile function $\mu(r)$.

In Table 3 the numerical results for the vacuum energy $\E_{\rm
quant}^{\rm renPV}$ given by Eq.\Ref{Etot} are shown. The dominating
part is the asymptotic contribution $\Eas$, followed by $\Easn$. The
finite contributions $\Ef$ and $\Efn$ are considerably smaller.

We conclude the representation of the numerical results by  Table 4
showing the quantities entering the calculation of the minimum of the
complete energy, namely $A$, Eq.\Ref{A}, $B$, Eq.\Ref{B}, the radius
${r_0}_{\mid_{\rm min}}$, Eq.\Ref{rmin} and the energy in the minimum,
$\E_{\mid_{\rm min}}$, Eq.\Ref{egesmin} (the latter two quantities are
given in units of ${\Lambda_{\rm QCD}}$).

\begin{table}[t]\label{tab2}
\begin{tabular}{|cccccc|}\hline&&&&&\\[-9pt]
N& $\Easn$ & $\Eas$ &$\Efn$&$\Ef$&$ \E_{\rm quant}^{\rm renPV}$
\\\hline  &&&&&\\[-5pt]
\multicolumn{6}{|c|}{$\mu(r)=\tanh^2r$} \\
2& -0.352761 &  0.560818 &  -0.0482214 &  -0.00659258 &  0.153243  \\
3&  -0.526636 &  0.83779 &  -0.0894967 &  0.00379912 &  0.273678
\\\hline  &&&&&\\[-5pt]
\multicolumn{6}{|c|}{$\mu(r)=\tanh^4r$} \\
2& -0.0848931 &  0.223159 &  -0.0889963 &  0.000655135 &  0.0499246 \\
3&  -0.126305 &  0.332705 &  -0.12675 &  0.0101892&   0.178836
\\\hline  &&&&&\\[-5pt]
\multicolumn{6}{|c|}{$\mu(r)=\tanh^8r$} \\
2&  -0.0424693 &  0.124992  & -0.0649512&   0.00630883 &  0.0238805  \\
3& -0.0631556  & 0.186119 &  -0.0903911 &  0.0161882 &  0.113711
\\\hline  &&&&&\\[-5pt]
\multicolumn{6}{|c|}{$\mu(r)=r^6/(1+r^6)$} \\
2&  -0.188802 &  0.499386 &  -0.08936&   0.0452825 &  0.266506 \\
3& -0.280762 &  0.745303  & -0.137446&   0.0877988 &  0.504254  \\\hline
\end{tabular}\caption{The constituents of the vacuum energy according
  to Eq.\Ref{Etot}.}
\end{table} 
%
\begin{table}[h]\label{tab3}
\begin{tabular}{|ccccc|}\hline&&&&\\[-9pt]
N& $A$ & $B$ &${r_0}_{\mid_{\rm min}}{\Lambda_{\rm QCD}}$&
$\E_{\mid_{\rm min}}/{\Lambda^2_{\rm QCD}} $\\\hline  &&&&\\[-5pt]
\multicolumn{5}{|c|}{$\mu(r)=\tanh^2r$} \\            
2& 0.864328& 0.148238& 11.1221& -0.000324477 \\
3& 0.864328& 0.103024& 16.0032& -0.000235092\\ \hline  &&&&\\[-5pt]
\multicolumn{5}{|c|}{$\mu(r)=\tanh^4r$} \\             
2& 0.488359& 0.332642& 4.95644& -0.000923169 \\
3& 0.488359& 0.072185& 22.8402& -0.0000652096 \\ \hline  &&&&\\[-5pt]
\multicolumn{5}{|c|}{$\mu(r)=\tanh^8r$} \\             
2& 0.353234& 0.482925& 3.41403& -0.00140737 \\
3& 0.353234& 0.0991947& 16.6211& -0.0000890673  \\ \hline  &&&&\\[-5pt]
\multicolumn{5}{|c|}{$\mu(r)=r^6/(1+r^6)$} \\           
2& 1.07484& 0.0692783& 23.7985& -0.0000881308 \\
3& 1.07484& 0.0344776& 47.8201& -0.0000327415\\ \hline
\end{tabular}\caption{Summary for the energy of the vortex}
\end{table}
 
\section{Conclusions}\label{sec6} 
We calculated the vacuum energy for a color magnetic vortex in QCD.
We found that the renormalized vacuum energy ${\E_{\rm quant}^{\rm
renPV}}$ calculated at $r_0=1$ is positive. As discussed in Sec. 2 the
sign of the complete energy in the minimum $\E_{\mid_{\rm min}}$,
Eq.\Ref{egesmin}, is not affected by this sign but only its deepness.
For a positive ${\E_{\rm quant}^{\rm renPV}}$ it is less deep. At once
the core radius becomes larger. In this sense $\mu(r)=\tanh^8r$ is the
'best' one among the considered profile functions.

In general, the sign of the vacuum energy is hard to predict.  In the
given case the vacuum energy ${\E_{\rm quant}^{\rm renPV}}_{\mid
r_0=1}$ which is positive results from bosonic fluctuations whereas in
the magnetic background calculated in
\cite{Bordag:1998tg,Groves:1999ks,Langfeld:2002vy} it is negative
resulting from fermionic ones. However, these results cannot be
compared directly because of different normalization conditions used
(massless vs. massive case).  Nevertheless, one may wonder to what
extend this is a general rule and what the quark contributions might
change for a center-of-group vortex.

The topic of stability of the vortex is to a large extend beyond the
scope of the present paper. So we only remark that the bound states
and the imaginary part found here make the whole configuration
unstable with respect to creation of particles. This is the same
instability as known from the homogenous color magnetic field. 

The bound states are well known in a homogeneous color magnetic
field. In an inhomogeneous one their existence can be predicted from
some simple arguments. First note that the background we are concerned
with is in fact Abelian and the contribution proportional to $\mu'(r)$
in Eq.\Ref{radeq} is an additional magnetic moment. From
\cite{AharonovCasherj} it is known that for a balance between the
lowest Landau level and an attractive interaction between the magnetic
moment and the magnetic background field there are zero modes, one for
each flux quantum. In \cite{AharonovCasherj} this had been established for an
electron but it is obvious that this holds in our case too. Here the
magnetic moment is larger due to the spin of the gluon. The zero modes
become more tightly bounded and turn into bound states. This is
independent from the specific form of the background and it is what we
observed for our profile functions $\mu(r)$. The bound states found
here are also in line with those in \cite{Bordag:1994ta} where an
electron with gyromagnetic factor larger than 2 had been considered.

The calculations had been performed in an efficient way, using Jost
functions and separating an asymptotic part thus leaving fast
convergent expressions. The main numerical effort is for the finite
parts. Here one has to perform the integration over the radial
momentum and the sum over the orbital momenta which are fast
converging due to the subtraction of the corresponding first few terms
of the asymptotic expansions. For instance, in Eq.\Ref{Ef1} we
subtract up to $O\left(\frac{1}{\nu^3}\right)$ which is sufficient for
convergence. It turns out that the next power, $\frac{1}{\nu^4}$ is
absent and we are left with an $\frac{1}{\nu^5}$-behavior making the
sum in Eq.\Ref{Ef1} fast convergent. The absence of the
$\frac{1}{\nu^4}$-contribution is due to cancellations between the
orbital momenta of different sign\footnote{We note that this had been
taken into account in the coefficients $X_{nj}$ and $Z_{nj}$ which are
listed in the Appendix.}. Furthermore, this cancellation is related to
the smoothness properties of the background. If the background has a
jump, as, for instance, considered in
\cite{Bordag:1998tg,Drozdov:2002um,Groves:1999ks}, a contribution
proportional to $\frac{1}{\nu^4}$ is present slowing down the
convergence. Of course, one may think about over subtractions, i.e.,
including more terms in the asymptotic part of the Jost function,
Eq.\Ref{Jexp}. But there one needs more orders in the iteration of the
Lippmann-Schwinger equation than done in \cite{Bordag:1998tg}. This
had been done only for the particular example of the pure magnetic
background in \cite{Drozdov:2002um}.

In Sec. 5 it was found that the numerically dominating contribution
results from $\Eas$. It is of some interest to discuss the question
whether $\Eas$ can be attributed to a orbital momentum. We start with
the remark that the initial, regularized vacuum energy $\Eq$, \Ref{eq2},
which appears after separation of variables, is a sum over orbital
momenta. This remains true also after the subdivision into finite and
asymptotic parts in Sec. 3 and for the finite parts themselves.  Also
$\Easn$ can be attributed to the s-wave. However, $\Eas$ cannot. Here
the reason is that in order to perform the analytic continuation to
$s=0$ we used the Abel-Plana formula and performed the integration
over $\nu$ after what it is impossible to identify contributions from
some $\nu$. In this way we cannot say from which orbital momentum the
numerically dominating contribution originates. To some extend this
discussion can be continued with the remark that it is the counter
term who cannot be  represented as a sum over orbital momenta.

\section*{Acknowledgments} The author is indebted to D. Diakonov for
directing his attention to this problem and for interesting
discussions during a very pleasant visit to Copenhagen. The author
thanks K. Kirsten, H. Gies, V. Skalozub and D. Vassilevich for
interesting and helpful discussions.
\section*{Appendix}\label{App}
The coefficients $X_{nj}$ appearing in Eq.\Ref{Jexp} read
\beao\label{25}
X_{1,1}&=&\frac{{\alpha_c}^2\,{\mu(r)}^2}{2}  , \\
X_{1,2}&=&0  , \\ 
X_{1,3}&=&\frac{-\left( {\alpha_c}^2\,{\mu(r)}^2 \right) }{2}, \\
X_{2,2}&=&\frac{{\alpha_c}^2\,{\mu(r)}^2}{2} - \
\frac{{\alpha_c}^2\,r\,\mu(r)\,\mu'(r)}{2}, \\
X_{2,3}&=&-\left( {\alpha_c}^2\,\beta_\la\,r\,\mu(r)\,\mu'(r) \right) , \\
X_{2,4}&=&\frac{-5\,{\alpha_c}^2\,{\mu(r)}^2}{2} + \
{\alpha_c}^2\,r\,\mu(r)\,\mu'(r), \\
X_{2,5}&=&0, \\
X_{2,6}&=&2\,{\alpha_c}^2\,{\mu(r)}^2, \\
X_{3,3}&=&\frac{13\,{\alpha_c}^2\,{\mu(r)}^2}{16} - \
\frac{{\alpha_c}^4\,{\mu(r)}^4}{8} - {\alpha_c}^2\,r\,\mu(r)\,\mu'(r) + \
\frac{{\alpha_c}^2\,r^2\,{\mu'(r)}^2}{4}  \\ && - \
\frac{{\alpha_c}^2\,{\beta_\la}^2\,r^2\,{\mu'(r)}^2}{2} + \
\frac{{\alpha_c}^2\,r^2\,\mu(r)\,\mu''(r)}{4}, \\
X_{3,4}&=&-3\,{\alpha_c}^2\,\beta_\la\,r\,\mu(r)\,\mu'(r) + \
{\alpha_c}^2\,\beta_\la\,r^2\,{\mu'(r)}^2 + \
{\alpha_c}^2\,\beta_\la\,r^2\,\mu(r)\,\mu''(r), \\
X_{3,5}&=&\frac{-153\,{\alpha_c}^2\,{\mu(r)}^2}{16} + \
\frac{3\,{\alpha_c}^4\,{\mu(r)}^4}{4} + \
\frac{27\,{\alpha_c}^2\,r\,\mu(r)\,\mu'(r)}{4} \\ && - \ 
\frac{5\,{\alpha_c}^2\,r^2\,{\mu'(r)}^2}{8} - \
\frac{3\,{\alpha_c}^2\,r^2\,\mu(r)\,\mu''(r)}{4}, \\
X_{3,6}&=&4\,{\alpha_c}^2\,\beta_\la\,r\,\mu(r)\,\mu'(r), \\
X_{3,7}&=&\frac{315\,{\alpha_c}^2\,{\mu(r)}^2}{16} - \
\frac{5\,{\alpha_c}^4\,{\mu(r)}^4}{8} - \
\frac{25\,{\alpha_c}^2\,r\,\mu(r)\,\mu'(r)}{4}, \\
X_{3,8}&=&0, \\
X_{3,9}&=&\frac{-175\,{\alpha_c}^2\,{\mu(r)}^2}{16}.
\eeao
The non vanishing coefficients $Z_{nj}$ appearing in Eq.\Ref{asZ} read
\bea\label{Znj}
Z_{1,1}&=&{\alpha}^2\,r\,\mu(r)\,\mu'(r)  ,\nn \\
Z_{2,3}&=&-\left( {\alpha}^2\,\beta\,r\,\mu(r)\,\mu'(r) \right)   , \nn \\
Z_{3,3}&=&\frac{-\left( {\alpha}^4\,r\,{\mu(r)}^3\,\mu'(r) \right) \
}{6} - \frac{{\alpha}^2\,\left( -1 + 2\,{\beta}^2 \right) \
\,r^2\,{\mu'(r)}^2}{4} \nn \\&& + \ \frac{{\alpha}^2\,r\,\mu(r)\,\left( \
-11\,\mu'(r) + 6\,r\,\mu''(r) \right) }{24}  ,\nn  \\
Z_{3,4}&=&{\alpha}^2\,\beta\,r^2\,{\mu'(r)}^2 + \
{\alpha}^2\,\beta\,r\,\mu(r)\,\left( -3\,\mu'(r) + r\,\mu''(r) \
\right)   ,\nn  \\
Z_{3,5}&=&\frac{{\alpha}^4\,r\,{\mu(r)}^3\,\mu'(r)}{2} - \
\frac{5\,{\alpha}^2\,r^2\,{\mu'(r)}^2}{8} - \
\frac{{\alpha}^2\,r\,\mu(r)\,\left( -13\,\mu'(r) + 3\,r\,\mu''(r) \
\right) }{4}  ,\nn  \\
Z_{3,6}&=&4\,{\alpha}^2\,\beta\,r\,\mu(r)\,\mu'(r)  ,\nn  \\
Z_{3,7}&=&\frac{-25\,{\alpha}^2\,r\,\mu(r)\,\mu'(r)}{8}  .
\eea
The Abel-Plana formula used in Sect.2 reads
\be\label{APla}
 \sum\limits_{\nu=1}^{\infty}f(\nu)=
\int\limits_{0}^{\infty}d \nu~f(\nu)
-\frac{1}{2}f(0)
+\int\limits_{0}^{\infty}{d \nu\over 1-e^{2\pi\nu}}{f(i\nu)-f(-i\nu)\over i}.
\ee
The following formulas are used in the text,
\be\label{C3}
\int_m^\infty dk(k^2-m^2)^{1-s}{\partial\over\partial  k}t^{j}
=
- m^{2-2s}
{\Gamma(2-s)\Gamma(s+\frac{j}{2}-1)\over  \Gamma(\frac{j}{2})}
 { \left({\nu\over mr }\right)^{j-n} \over
   \left(1+\left({\nu\over mr}\right)^{2}\right)^{s+\frac{j}{2}-1}}
\ee
and
\be\label{C2}
\int_{0}^{\infty}d\nu\int_{m}^{\infty}dk(k^{2}-m^2)^{1-s}{\partial\over\partial
  k}{t^{j}\over\nu^{n}}
=
-{m^{2-2s} \over   2(rm)^{n-1}}{\Gamma(2-s)\Gamma(\frac{1+j-n}{2})
\Gamma(s+\frac{n-3}{2})\over\Gamma(j/2)}
\ee
with $t=1/\sqrt{1+(kr/\nu)^2}$.  They can be easily derived, see also
(C3) and (C2) in \cite{Bordag:1998tg}.

\bibliographystyle{unsrt} 
\bibliography{/home/bordagm/Literatur/Bordag,/home/bordagm/Literatur/articoli,/home/bordagm/Literatur/libri,/home/bordagm/Literatur/Kirsten}
\end{document}